\documentstyle[preprint,aps]{revtex} 
\begin{document}
\draft
\begin{title}
{Phase Transition in the Two Dimensional Classical XY Model}
\end{title}
\author{Jae-Kwon Kim}
\address
{Center for Simulational Physics and Department of Physics,\\
The University of Georgia, Athens, GA 30602}
\maketitle

\begin{abstract}
For the two dimensional classical XY model 
we present extensive 
high -temperature -phase bulk data extracted based 
on a novel finite size scaling (FSS) Monte Carlo technique, 
along with FSS data near criticality. 
Our data verify that $\eta=1/4$ sets in 
near criticality, and clarify the nature of correction to the 
leading scaling behavior. 
However, the result of standard FSS
analysis near criticality is inconsistent with 
other predictions of Kosterlitz's renormalization group approach. 
\end{abstract}
\pacs{PACS Numbers: 05.70.Fh, 64.60.Fr, 75.10.Hk, 75.40.Cx}

The classical XY model is defined by the Hamiltonian
\begin{equation}
H = -{1 \over T} \sum_{<ij>} {\vec{s}_{i} \cdot \vec{s}_{j}} \label{eq:ham},
\end{equation}
where $\vec{s}_{i}$ is the O(2) spin at site $i$ and the sum is
over nearest-neighbor spins.
As a novel statistical model that undergoes 
a phase transition without long range order,
the two dimensional (2D) XY model is interesting in its own right,
as well as being a model for 2D layers of either
superconducting materials or films of liquid helium.

The mechanism of this kind of phase transition was  first illustrated by
Berezinskii\cite{BER}, and by Kosterlitz and Thouless\cite{TOU}(BKT)
based on vortex binding scenario: 
At low-temperatures
spin waves are the only significant fluctuations, and 
vortices are bound in pairs of zero vortices, thus not affecting  
the spin-wave description qualitatively; at high temperatures, however,
the binding of the vortices decreases.
The prediction\cite{KOS} obtained with a
renormalization-group method is that in the high temperature phase
the critical properties of correlation length ($\xi$) and 
magnetic susceptibility ($\chi$) are given by
\begin{eqnarray} 
\xi (t) &\sim& \exp (b t^{-\nu}), \label{eq:ktcor} \\
\chi (t) &\sim& \xi(t)^{2-\eta}. \label{eq:ktsuc}
\end{eqnarray}
Here $t= (T-T_{c})/T_{c}$, and  
the predicted values of $\nu$ and $\eta$
are 1/2 and 1/4, respectively.
For  $T \le T_{c}$, both $\xi$ and $\chi$ diverge identically with
such a temperature dependent $\eta$ that $\lim_{T \to 0} \eta(T)=0$.

Kosterlitz's renormalization group equations approach was
extended to yield 
the corrections to Eq.(\ref{eq:ktcor}):
\begin{eqnarray} \label{eq:cor}
\xi (t)  \sim \exp (b t^{-1/2})~ [1 + {\cal O}(t)],~~~
\chi (t) \sim t^{-1/16} \exp ({7 \over 4} b t^{-1/2}) 
\end{eqnarray}
Accordingly $\chi (t)/\xi (t)^{7/4}$ increases as $t \to 0$.

On the other hand, rigorous studies have established
only two things: the existence of a phase transition
at a finite temperature\cite{FRO} and also the existence
of an  upper bound of the critical temperature\cite{SIM}. 
Consequently the validity of Eq.(\ref{eq:ktcor}) 
has been questioned both analytically\cite{LUT} and 
numerically\cite{PAT,SEI,SOK}.
A different mechanism, namely ``polymerization of the domain walls", 
was proposed\cite{SEI}.
Extensive Monte Carlo (MC) studies \cite{WOL,GUP,JAN}
up to $\xi \simeq 70(1)$ and series expansions studies 
\cite{BUT,CAMP} supported Eq.(2) over a {\it pure} 
power-law singularity. But  neither the MC data nor the 
series expansions showed $\eta=0.25$ and $\nu=0.5$ conclusively.

Primary difficulties of various numerical methods 
in locating the parameters in Eq.(\ref{eq:ktcor}) 
are due to four independent
parameters involved so that, for example, a determination
of $\nu$ is sharply sensitive to the location of $T_{c}$,
and vice versa. 
Thus, most MC studies have determined $T_{c}$ by assuming $\nu=1/2$ in
Eq.(\ref{eq:ktcor}), yielding 
$ 0.890 \le T_{c} \le  0.894$
\cite{WOL,GUP,JAN,BUT,CAMP,CHE,FER,OLS,HAS}.
The generic feature that has emerged from those MC studies
was that the values of $\eta$ calculated from the 
thermodynamic values (TV value in the thermodynamic limit) of 
$\xi$ and $\chi$ are significantly larger 
than the predicted value\cite{GUP,JAN,KIME}.

In this work we present extensive MC data of the thermodynamic
values of $\xi$ and $\chi$ up to T=0.93 ($\xi \simeq 1391(22)$),
indicating for the first time that $\eta=1/4$ holds as $T \to T_{c}$. 
The {\it unconstrained} fit of our thermodynamic data are
remarkably consistent with the prediction of Kosterlitz,
yielding $\nu \simeq 0.48$ and $T_{c} \simeq 0.893$.
The nature of the corrections to the leading scaling behavior 
Eq.(\ref{eq:ktsuc}) is clarified.
We also attempt to locate $T_{c}$ by conventional FSS
, i.e., by finding
the temperature where $\chi_{L} \sim L ^{2-\eta}$ holds 
with $\eta=1/4$ and, at the same time, where 
the 4th order cumulant ratio \cite{BIN} is invariant with $L$. 
This procedure yields $T_{c}$ over the 
range $0.900 < T_{c} <0.905$, being incompatible with
that estimated by assuming Eq.(2).

Our extractions of TVs are based
on a novel FSS technique\cite{KIME,KIMD,SOZ,SOK2} that facilitates 
drastically the MC measurements of TV. 
Namely, thermodynamic data can be computed from
MC measurements on a much smaller lattice than 
required for the traditional direct measurements.
The technique is based on the fundamental formula 
of FSS\cite{KIME},
\begin{equation}
A_{L}(t)=A_{\infty}(t) {\cal Q}_{A}(x),~~~x=\xi_{L}(t)/L. \label{eq:fss}
\end{equation}
Here $A_{L}$ denotes a multiplicative renormalizable 
quantity $A$, defined on a finite lattice 
of linear size L, and ${\cal Q}_{A}$ is a universal
scaling function of a different scaling variable $\xi_{L}(t)/L$.
Eq.(\ref{eq:fss}) is supposed to hold even for 
the 2D XY model\cite{BRE}.
For a detailed description of the technique  
see Refs.\cite{KIME,KIMD,SOZ},
where the TVs extracted by employing the technique
are shown to agree completely with those measured traditionally
for the 2D and 3D Ising models\cite{SOZ}, 2D XY model\cite{KIME},
and 2D Heisenberg ferromagnet\cite{KIMD,SOK2}.

Standard theory of FSS predicts the presence of $L_{min}$
below which certain corrections to Eq.(\ref{eq:fss}) 
become non-negligible.
It is shown for the 2D and 3D Ising models\cite{SOZ} that 
at least for $L \ge 16$ Eq.(\ref{eq:fss}) holds
(hence, $L_{min} \le 16$) within 
the accuracy of the relative statistical errors of 0.3 percent.
For  $T$ not so close to the criticality in the high
temperature phase of the XY model, Eq.(\ref{eq:fss}) holds
with $L_{min} \simeq 30$\cite{KIME}.
As will be seen, however, $L_{min}$ increases as $T \to T_{c}$.
Whether or not  $L_{min}$ will increase
endlessly as $T\to T_{c}$ (thus violating FSS) is not clear at present, 
although its violation seems unlikely on theoretical ground\cite{BRE}. 
We will assume that Eq.(\ref{eq:fss}) is accurate 
for any value of $t$ provided $L \ge 80$\cite{COM1}.

Here the extractions of the TVs of $\xi$ and $\chi$ 
are made in the range $0.93 \le T \le 0.98$,  
based on available data of $(x, {\cal Q}_{A}(x))$ 
for $T=1.0$ \cite{KIME}.
The values of L over the range $80 \le L \le 480$ are used for
our analysis, and most interpolations are made in the 
range ${\cal Q}_{A}(x) \ge 0.1$.
We employed Wolff's single- cluster algorithm\cite{WOL1} 
for the MC simulations, imposing periodic 
boundary condition on a square lattice.
For a given  set of $T$ and L, typically 20 - 40 numbers of bins, 
(5 - 10  bins for the conventional FSS studies near criticality 
for the location of $T_{c}$), have been obtained for 
the measurements.  Each bin is
composed of 10 000 measurements, each of which
is separated by 3-12 consecutive one cluster updating.  
Accordingly, the relative statistical errors of most raw-data
calculated through the jack knife method are  typically 
less than 0.1 percent for the $\chi$ and 0.2 percent for the $\xi$.
Correlation lengths are measured by the low-momentum behavior
of the propagator\cite{PAT,GUP,KIME,KIMD,SOZ,SOK2}.
A comparison of our thermodynamic data in Table(1) with those
obtained from a strong coupling analysis can be found in
Ref.\cite{CAMP}, which shows  good agreement.

Based on the bulk data, ${\cal Q}_{A}(x)$  for $A=\xi$ and 
$\chi$ are  plotted in Fig.(1), displaying
an excellent data-collapse for $L \ge 80$. The data-collapse verifies
Eq.(\ref{eq:fss}) for the given values of L and $T$.
The figure also indicates a deviation from the data -collapse for 
the data point corresponding to $T=0.95$ and $L=60$. Similar 
figures including data points with $L \le 60$
will be presented in a detailed paper\cite{LAN2}. 
They will serve for a clear demonstration for the effect of 
correction to FSS for these values of L. 
Fig.(2) depicts $\chi(T)/\xi(T)^{2-\eta}$ with $\eta=1/4$
as a function of $\ln(\xi)$. It shows
that the effective value of $\eta$ gradually decreases 
to $\eta=1/4$ as $T \to T_{c}$. 
Empirical formula obtained from the data in Fig.(2), 
assuming $T_{c} = 0.893$ (see below), is as follows:
\begin{equation}
\chi(t)/\xi(t)^{7/4} = a +b |\ln (t)|^{r} \label{eq:eta_cor},
\end{equation}
where $a \simeq 1.833$, $b=0.955$, and $r \simeq -0.413$.
Note that the slow decrease of 
$\chi(t)/\xi(t)^{7/4}$ as $t \to 0$ is inconsistent with 
any positive value of $r$ such as was predicted in \cite{AMI}.
Without taking account of the corrections to Eq.(\ref{eq:ktsuc}),
it turns out, however, that $\eta \simeq 0.272(8)$.
The value of $\chi^{2}$ per degree of freedom ($\chi^{2}/N_{DF}$) 
for the latter case turns out to be approximately $5.4 \times 10^{2}$,
which is much larger than that for Eq.(\ref{eq:eta_cor}). 
Accordingly, correction to 
Eq.(\ref{eq:ktsuc}) is essential.

We also employ the conventional FSS technique to 
locate $T_{c}$ independently.  Namely,
we measured both $\chi_{L}$ and the 4th order cumulant ratio 
defined as $U_{L}=3- <S^{4}>/<S^{2}>^{2}$ with
$S^{2}\equiv |\sum_{i}\vec{s}_{i}|^{2}$.
Such well-known standard FSS properties 
at criticality as the scale invariance of the $U_{L}$ and 
$\chi_{L} \sim L^{2-\eta}$ are direct consequences of
the fundamental FSS formula, Eq.(\ref{eq:fss}).
Therefore those properties should be valid 
only for $L \ge L_{min}$.
We would like to stress, moreover, that 
an additional criterion should be satisfied in order for
the former property to be true:
a weak hyperscaling relation that
the renormalized four point coupling at zero momentum ($g_{R}^{(4)}$) 
remains a constant in the scaling region\cite{COM2}.
Such a scaling behavior of $g_{R}^{(4)}$ has recently been 
shown numerically\cite{KIMB} for the 2D XY and 
$O(3)$ vector models.

We measured $U_{L}$ and $\chi_{L}$ by varying L from 20 to 600
for some temperatures over the range $0.89 \le T \le 0.92$.
Certain {\it non-asymptotic} FSS behavior was observed for
$L < 80 $\cite{LAN2}, which seems to be a consequence of 
the correction to Eq.(\ref{eq:fss})\cite{COM3}. 
Our data with $L \ge 80$ fit remarkably well to a standard FSS 
formula at the criticality
\begin{equation}
\chi_{L} \sim L^{2-\eta}  \label{eq:chi}
\end{equation}
for $T \le 0.905$ ($\chi^{2}/N_{DF} < 0.5$) (Fig.(3)).
The extracted values of $\eta$ from the linear fit are:
$\eta(T)=0.260(1)$, 0.252(0), 0.245(0), and 0.231(1) for
T=0.91, 0.905, 0.90, and 0.89 respectively.
Thus $\eta (T_{c})=0.25$ is consistent with 
$0.90 \le T_{c} \le 0.905$, most probably $T_{c}\simeq 0.904$.
This estimate is also consistent with our data of 
$U_{L}$ (Fig.(4)).  $U_{L}(T=0.92)$ definitely
decreases  with L from L=80 to L=240, so that 
$T_{c}$ is obviously smaller than 0.92.
At T=0.91 and 0.905 it varies so mildly that 
some fictitious scale invariance of $U_{L}$ 
appears over a range of L not too large;
nevertheless, it eventually decreases.
This fictitious scale invariance may result from an extremely 
large value of $\xi$ over the range of temperature. 
We observe the invariance of
$U_{L}(T)$ with respect to L at $T=0.90$, 
indicating $T_{c} \ge 0.90$.

We carried out a $\chi^{2}$ fit of $\xi$ data over 
the range  $5.01(3)  \le \xi \le 1391(22)$ (corresponding to
$0.93 \le T \le 1.19$) to the exponential singularity, 
Eq.(\ref{eq:ktcor}). Our fit, based on down-hill simplex
algorithm, is highly non-linear for Eq.(\ref{eq:ktcor}), so that
we have treated $T_{c}$ as an input parameter over a reasonable range.
The values of $\chi^{2}/N_{DF}$ and $\nu$ as a function of the
input $T_{c}$ are plotted in Fig.(5). 
The best fit is obtained for $T_{c}\simeq 0.893$
and for $\nu \simeq 0.48$ ($\chi^{2}/N_{DF} \simeq 0.75$).
$\nu=0.5$ holds at $T_{c} \simeq 0.892$
with $\chi^{2}/N_{DF} \simeq 0.80$, being in agreement 
with the range of $T_{c}$ conventionally accepted.
The agreement is gratifying and may be regarded as another 
verification for Eq.(\ref{eq:fss}) with $L_{min}=80$, 
at least for $T \ge 0.93$.
The results of the unconstrained fit 
agree extremely well with the prediction in Eq.(2).
Notice, however, that $\eta=1/4$ holds for $T_{c} \simeq 0.904$
according to the conventional FSS, being much larger than 0.892.
Assuming $T_{c}=0.904$ the best fit is for $\nu \simeq 0.30(1)$ 
with $\chi^{2}/N_{DF} \simeq 6.69$; 
assuming both $T_{c}=0.904$ and $\nu=0.5$ the best fit has
$\chi^{2}/N_{DF} \simeq 263$. Therefore,
$\nu=1/2$ in Eq.(2) is definitely inconsistent with 
the prediction of FSS at criticality.

We also checked whether or not the data fit to 
$\xi=c_{1}(1+c_{2} t^{\theta}) \exp (bt^{-\nu})$
with the input value of $T_{c}$ over $0.90 \le T_{c} \le 0.905$.
It turns out that they fit to the formula 
for some {\it negative} value of $\theta$ only 
(being inconsistent with Eq.(4)),
with the value of $\chi^{2}/N_{DF}$ being almost insensitive 
to the choice of $T_{c}$. However,
this function has too much freedom to the fit
and gives unstable predictions for the values of both 
$T_{c}$ and $\nu$.

On the other hand, 
the  data fit very well to a modified 
second order phase transition (see also Ref.\cite{SOK}), 
\begin{equation}
\xi =c_{1} (1+ c_{2} t^{\theta})t^{-\nu}.  \label{eq:mpl}
\end{equation}
Assuming $T_{c}=0.904$ we estimate that
$\nu \simeq 3.10(8)$, $\theta \simeq 1.82(8)$, 
and $c_{2} \simeq 44.2$ 
with $\chi^{2}/N_{DF} \simeq 1.2$ (Fig.(6)). 
As the value of the fixed $T_{c}$ becomes larger,
the data fit better to the modified power-law singularity. 
The large value of $c_{2}$ is unusual, implying that
at sufficiently large $t$ the correction term becomes dominant
with an effective exponent of $\xi$ being equal to $\nu -\theta$.
We note that our effective $\nu$ for large $t$
agrees with the prediction in \cite{JON}.
For  $t$ small enough the correction becomes negligible
regardless of the value of $c_{2}$, while
over the intermediate $t$ the effective critical exponent
changes gradually. 

Assuming $T_{c} \simeq 0.904$,
even a modified exponential singularity including 
correction is unlikely:
For a power-law singularity, $\xi \sim t^{-\nu}$, the FSS of 
$U_{L}(t)$ is given by, $U_{L}(t) \sim f_{U}(L^{1/\nu}t)$ \cite{BIN}.
Using the analyticity of $U_{L}$ for a finite L, after a Taylor
expansion we get $U_{L}(t) \simeq U_{L}(0)+ c L^{1/\nu}t + 
{\cal O}((L^{1/\nu}t)^{2})$ for a sufficiently small 
value of the $L^{1/\nu}t$.
An exponential singularity corresponds to 
$\nu \to \infty$ so that
$U_{L}(T)$ at $T=0.905$ ($t \simeq 10^{-3}$), for example,
would be L independent regardless of the value of L,
contrary to our finding in Fig.(4).
Similarly, if an exponential singularity is  valid
$\chi_{L}(T) \sim L^{2-\eta (T_{c})}$ would hold 
identically at any $T$ close to $T_{c}$; 
in other words, $\eta$ would remain unchanged
over a finite range of $T$ close to $T_{c}$,  
which again contradicts our data in Fig.(3). 
In view of this FSS argument, an exponential singularity
could be consistent with our data of $\chi_{L}$ (Fig.(3))
and $U_{L}$ (Fig.(4)) only if $T_{c}$ is considerably 
smaller than 0.904. 
Assuming a power-law singularity, on the other hand, 
one can easily check from the expansion of $\chi_{L}(t)$ that
even for such a modestly large value of $\nu$ as $\nu \simeq 3.1$,
$\ln (\chi_{L}/L^{7/4})$ at T=0.905 varies very mildly with L
in agreement with our data in Fig.(2)\cite{LAN2}.

It remains to be seen, however, whether or not $L_{min}$
diverges as $T \to T_{c}$. In this case, the conventional  
FSS analysis at the criticality would be misleading.
It is also possible that the correction to the leading
scaling behavior (Eq.(\ref{eq:ktsuc})) results in a modification
to the FSS behavior (at criticality). 
The conventional analysis then would not
give a very accurate estimate of $\eta(T_{c})$.
A more detailed account for the possible failure of the 
standard FSS analysis at the criticality, along with details of
our analysis and methods of obtaining our data, 
will be presented in a longer paper\cite{LAN2}.

In conclusion we have reported strong numerical evidence that
$\eta=1/4$ sets in at temperatures close to the criticality. 
The nature of the correction to the leading scaling behavior
Eq.(\ref{eq:ktsuc}) has been clarified. 
The {\it unconstrained} fit of our high temperature 
thermodynamic data yields results consistent with
the predictions of the BKT for the values of $\nu$ and $\eta$.
This, nevertheless, does not rule out the possibility of
the ordinary second order phase transition. 
Knowing the {\it true} critical point is crucial to the resolution
of the order of the phase transition in the 2D classical XY model.
The critical point obtained from the conventional FSS analysis, 
however, seems to be
incompatible with the conventional exponential singularity. 

After the completion of the current work we have become
aware of a recent related work by Kenna and Irving \cite{KEN} that 
questions the vortex binding scenario in the 2D XY model 
based on FSS study of Lee-Yang zeros. 
They argue that Eq.(2) cannot be compatible 
with $\eta=1/4$ without a certain correction. 
However, the proposed correction, $r \simeq 0.02$ 
in Eq.(\ref{eq:eta_cor}),
is {\it inconsistent} with our picture in Fig.(2) in 
that their $\chi(t)/\xi^{7/4}(t)$ {\it increases}  as $t \to 0$.

The author would like to thank Prof. M. Howard Lee for his
proof reading of this manuscript.
Current work was partially supported
by NSF grant DMR-9405018. 

\begin{table}
\caption{The thermodynamic values of $\xi $ and $\chi$ for 
        $0.93 \le T \le 0.98$, extracted through our MC FSS technique.}
\begin{tabular}{ccccccccc} 
T &0.98 &0.97 &0.96  &0.95 &0.945   &0.94  &0.935 &0.93\\
$\xi$ &70.4(4) &100.3(7) &155.7(1.5) &262.7(2.9) &364.5(3.6)
      &539.3(4.5)  &846.7(6.4) &1391(22) \\	
$\chi$ &4284(20) &7932(30) &16978(55) &42295(108) 
	&74742(206)   &147536(614) &3.24(1)$\times 10^{5}$
        &7.70(9)$\times 10^{5}$\\
\end{tabular}
\end{table}

{\bf Figure Captions: \\}
{\bf Fig.(1)}: ${\cal Q}_{\xi}(x)$ (upper data) and 
        ${\cal Q}_{\chi}(x)$ (lower data) for $L \ge 60$, 
        calculated with the data from Table(1). It should be clear that
	for each $T$ a data point with a larger L corresponds to
	a smaller $x$. For $T=0.95$ and L=60, the deviation from
	data-collapse is apparent. For the sake of the higher
	resolution of the data, we show only a lower part of 
        ${\cal Q}_{\xi}(x)$; for the upper part, see Ref.\cite{KIME}.\\
{\bf Fig.(2)}: $\chi (T)/\xi^{7/4}(T)$ versus $\ln(\xi (T))$ for T 
	 over the range, $0.93 \le
	 T \le 1.25$. The thermodynamic data for $T >0.98$ are 
         taken from Ref.\cite{KIME}, which agree completely with the 
	 data in \cite{WOL}\cite{GUP}. The tendency to decrease implies
	 $\eta > 0.25$, but the values of $\chi/\xi^{7/4}$ 
	 tend to stablize at approximately $\ln(\xi) \ge 5$.\\ 
{\bf Fig.(3)}: $\ln(\chi_{L}/L^{7/4})$ versus $\ln(L)$ over $80 \le L \le 600$.
	 Here the slope of a straight line is related to the value of
	 $\eta$, that is, $\eta=1/4-slope$. Note a slow decrease (increase)
 	 of $\ln (\chi_{L}/L^{7/4})$ with respect to $\ln (L)$  for
	 T=0.905 (T=0.90); hence, $0.90 < T_{c} < 0.905$ with $\eta=1/4$.\\
{\bf Fig.(4)}: $U_{L}$ as a function of $\ln (L)$ over $80 \le L \le 600$, 
         for various temperatures near criticality.
         $U_{L}$ is L independent (within small
	 statistical error) over $80 \le L \le 150$, and
	 $80 \le L \le 360$ for $T=0.91$ and 0.905 respectively (pseudo
         scale invariance), although it eventually decreases with L.  \\
{\bf Fig.(5)}: The values of $\chi^{2}/N_{DF}$ (lower) and $\nu$ (upper)
         as a function of the input parameter $T_{c}$, which are obtained 
         from the $\chi^{2}$
         fit of the $\xi$ data to Eq.(\ref{eq:ktcor}). \\
{\bf Fig.(6)}: $\ln (\xi)$ as a function of $|\ln t|$ assuming $T_{c}=0.904$. 
	 The circle symbols
         denote our data in Tab.(1), while the smooth curve
	 represents the modified second order phase transition,
	 Eq.(\ref{eq:mpl}), with the values of the parameters 
	 reported in the text (except for $c_{1} \simeq 2.19 \times 10^{-2}$).
	 The error bars are almost invisible.
	 Here a straight line would represent a pure power-law 
	 singularity, with its slope equivalent to the value of $\nu$.
	 Note that for $|\ln t| \ge 2.6$ the data are almost linear,
	 implying that the effect of correction is already not so
	 important in this regime.

\begin{references}
\bibitem{COM1} The violation of Eq.(\ref{eq:fss}) can be easily checked
	       by the observation of a deviation from data-collapse for 
	       $L < L_{min}$ when $(x(t),{\cal Q}_{A}(x(t))$ is plotted
               at different value of $t$. In this manner, it is shown 
               in Fig.(1) that $L_{min}=80$ is good enough for 
               $T \ge 0.93$. Nevertheless, a similar check for FSS
               at a temperature much closer to criticality seems 
               infeasible because accurate TVs are much difficult to 
               be measured.
\bibitem{COM2} $g_{R}^{(4)}$ has a well defined scaling behavior,
               $g_{R}^{(4)} \sim t^{D\nu-2\Delta+\gamma}$, while
               $U_{L}$ does not. Therefore, the FSS behavior of $U_{L}$
               should be derived from the FSS of $g_{R}^{(4)}$.
               From the definition of the $g_{R}^{(4)}$, and
               with an assumption of translational symmetry,
               it is trivial  to obtain
               $g_{R}^{(4)}(L,t)= (U_{L}(t)+C) (L/\xi_{L}(t))^{D}$
               where $C$ is a L independent constant.
               It should be clear from
               this relation that the scale invariance of $U_{L}$
               at criticality is a consequence of the vanishing of
               the critical index of $g_{R}^{(4)}$.
\end{references}
\end{document}